\documentclass[aps,prl,twocolumn,showpacs]{revtex4}

\usepackage{graphicx}
\usepackage{amsmath,amsfonts,amssymb}		
\usepackage{mathrsfs,bbm}

\def\aDOmegaX{\hat{a}^\dagger_{\mathrm{x},\omega}}
\def\aHOmegaX{\hat{a}^{\phantom{\dagger}}_{\mathrm{x},\omega}}

\def\aHOmegaY{\hat{a}^{\phantom{\dagger}}_{\mathrm{y},\omega}}

\newcommand{\qed}{\nobreak \ifvmode \relax \else
      \ifdim\lastskip<1.5em \hskip-\lastskip
      \hskip1.5em plus0em minus0.5em \fi \nobreak
      \vrule height0.75em width0.5em depth0.25em\fi}

\def\ket#1{|#1\rangle}

\def\Fz{F_\mathrm{z}}

\begin{document}

\title{Magnetic Field Estimation at and beyond 1/$N$ Scaling via an Effective Nonlinearity}
\author{Bradley A.~Chase}
\author{Heather L.~Partner}
\author{Brigette D.~Black}
\author{Benjamin Q.~Baragiola}
\author{Robert L.~Cook}
\author{JM Geremia}
\email{jgeremia@unm.edu}
\affiliation{Department of Physics \& Astronomy, The University of New Mexico, Albuquerque, New Mexico USA}

\begin{abstract}
We provide evidence, based on direct simulation of the quantum Fisher information, that $1/N$ scaling of the sensitivity with the number of atoms $N$ in an atomic magnetometer can be surpassed by double-passing a far-detuned laser through the atomic system during Larmor precession.   Furthermore, we predict that for $N \gg 1$, the proposed double-pass atomic magnetometer can essentially achieve $1/N$ scaling without requiring any appreciable amount of entanglement.
\end{abstract}
\date{\today}
\maketitle

\noindent \textit{Introduction---} The strength of a magnetic field $B$ is often determined by observing Larmor precession in a spin-polarized atomic vapor \cite{Budker:2002a,Romalis:2003a}.  More generally, a system of $N$ atoms can be prepared into a known initial state $\hat{\rho}_0$ and let to evolve to the final state $\hat{\rho}_\tau(B) = \hat{U}_\tau(B) \hat{\rho_0} \hat{U}_\tau^\dagger(B)$ at time $t=\tau$ under dynamics $\hat{U}_t(B)$ parameterized by the field $B$.   After (or preferably during) the evolution, the atoms can be measured to infer the field strength from the dynamics using the methods of quantum parameter estimation theory \cite{Wineland:1994a,Braunstein:1994a,Geremia:2003a,Chase:2008a}.

For precise measurements, the uncertainty $\delta \tilde{B}$ in the estimated value of the field strength $\tilde{B}$ is dominated by quantum fluctuations in the measurements performed on the atoms.  The quantum Cram\'{e}r-Rao inequality \cite{Helstrom:1976a,Braunstein:1994a} places an information-theoretic lower bound on the (units-corrected mean-square) uncertainty in terms of the quantum Fisher information $\mathcal{I}_t(B)$
\begin{equation} \label{Equation::CRError}
    \delta \tilde{B}_\tau = \left< 
     \left( \frac{\tilde{B}_\tau}{ |d \langle \tilde{B}_\tau \rangle / dB | } - B \right)^2
    \right>^\frac{1}{2} \ge \frac{1}{\sqrt{\mathcal{I}_\tau(B)}}, 
\end{equation}
which holds for any estimator used to determine $B$.  For pure states, the quantum Fisher information is given by the expectation value $\mathcal{I}_t(B) = \mathrm{tr} [ \hat{\mathfrak{L}}_t^2(B) \hat{\rho}_t(B) ]$ of the symmetric logarithmic derivative operator
\begin{equation}
	 \hat{\mathfrak{L}}_t(B) = 2  \partial \rho_t(B) / \partial B,
\end{equation}
which characterizes the sensitivity of the time-evolved state $\hat{\rho}_t(B)$ to variations in the value of the parameter $B$.   Subsequently, $\hat{\mathfrak{L}}_t(B)$ can be related to the generator of displacements in the parameter \cite{Braunstein:1994a,Boixo:2007a}
\begin{equation}
	\hat{\mathfrak{L}}_t = -2 i [ \hat{D}_t, \hat{\rho}_t ],  \quad
	\hat{D}_t(B) = \frac{\partial \hat{U}_t(B)}{\partial B} \hat{U}_t^\dagger(B).
\end{equation}

The particular value of the Fisher information (and therefore $\delta \tilde{B}$) achieved in a given setting depends upon both the initial state of the atomic system $\hat{\rho}_0$ and the nature of the dynamical evolution $\hat{U}_t(B)$ \cite{Boixo:2007a}.   When analyzing magnetometric limits, one typically takes  the dynamics to be generated by the  Zeeman Hamiltonian
\begin{equation} \label{Equation::HLarmor}
	\hat{H}(B) = 
	 - \hbar \gamma B  \,  \vec{n} \cdot \hat{\mathbf{F}},
\end{equation}
where $\vec{n}$ is the field unit vector $\mathbf{B} = B \vec{n}$, $\gamma$ is the atomic gyromagnetic ratio and $\hat{F}_\alpha =\sum_{j=1}^N \hat{f}_\alpha^{(j)}$ are the collective spin operators obtained from a symmetric sum over $N$ identical spin-$f$ atoms  ($F = f N$ for a sample of $N$ atoms each with total spin quantum number $f$).   The time-evolution operator then satisfies $i \hbar \partial_t \hat{U}_t(B) = \hat{H}(B) \hat{U}_t(B)$ and optimizing the variance $\langle \Delta^2 \hat{D}_\tau(B)\rangle$ over $\hat{\rho}_0$ for Eq.\ (\ref{Equation::HLarmor}) yields the shotnoise scaling
\begin{equation} \label{Equation::DeltaBSN}
	\delta \tilde{B}_\tau = 1/  \gamma \tau \sqrt{2 F} 
\end{equation}
when $\hat{\rho}_0$ is separable, and the Heisenberg scaling
\begin{equation} \label{Equation::DeltaBHL}
	\delta \tilde{B}_\tau = 1 /  \gamma \tau 2 F 
\end{equation}
when entanglement is permitted between the different atoms.  For $N$ spin-1/2 particles prepared into the initial cat-state $( \ket{\!\uparrow\uparrow \cdots \uparrow} + \ket{\!\downarrow\downarrow\cdots\downarrow})/\sqrt{2}$ (in a basis set by $\vec{n}$), the uncertainty scaling under Eq.\ (\ref{Equation::HLarmor}) is given by $1/\gamma \tau N$ and often called the \textit{Heisenberg Limit}.

\begin{figure}[t]
\vspace{2mm}
\begin{center}
\includegraphics{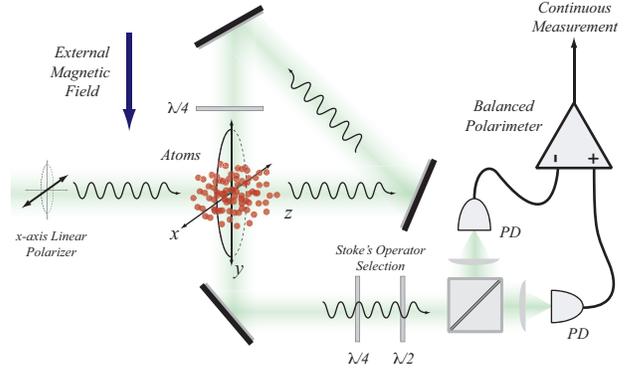}
\end{center}
\vspace{-6mm}
\caption{(color online) Schematic of a broadband atomic magnetometer based on continuous observation of a polarized optical field double-passed through the atomic sample.  \label{Figure::Schematic}}
\vspace{-3mm}
\end{figure}

It was believed for some time that Eqs.\ (\ref{Equation::DeltaBSN}) and (\ref{Equation::DeltaBHL}) were fundamental: the $1/\sqrt{N}$ scaling characteristic of the shotnoise uncertainty was an unavoidable byproduct of the coherent state projection noise $\langle \Delta \Fz\rangle =\sqrt{F/2}$, only improved via entanglement; and $1/N$ scaling could only be achieved by creating a cat-state or maximally squeezed state, but never surpassed \cite{Wineland:1994a}.  Recently, however, it was shown that the $1/N$ ``limit'' can be overcome \cite{Boixo:2007a} by extending the linear coupling that underlies Eq.\ (\ref{Equation::HLarmor}) to allow for multi-body collective interactions \cite{Boixo:2007a,Rey:2007a}.  Were one to engineer a probe Hamiltonian where $B$ multiplies $k$-body probe operators, such as $\hat{F}_\mathrm{y}^k$, then the optimal estimation uncertainty would scale more favorably, as $\Delta B_k \sim 1/N^k$ \cite{Boixo:2007a}.  Unfortunately, the Zeeman Hamiltonian is linear, meaning that $1/N$ scaling is optimal for magnetometry based solely on Eq.\ (\ref{Equation::HLarmor}).  Improving magnetometry beyond $1/N$ scaling therefore requires developing some kind of nonlinear magnetic interaction.

In this paper we provide evidence that one can improve upon the convenentional scalings in atomic magnetometry, Eqs.\ (\ref{Equation::DeltaBSN}) and (\ref{Equation::DeltaBHL}), by engineering effective dynamics that mimic a nonlinear coupling to the magnetic field using coherent positive feedback \footnote{Phenomena such as long-distance dipole-dipole interactions \cite{Ledbetter:2005a} or other collective effects might provide a nonlinearity suitable for improved magnetometry, however, our approach requires only minimal modification to current procedures based on Faraday spectroscopy.}.   Our approach involves double-passing an optical field through the atomic sample \cite{Sherson:2006a,Muschik:2006a,Sarma:2008a,Chase:2009a} at the same time that it is subject to Larmor precession, as depicted in Fig.\ (\ref{Figure::Schematic}).   We support our claim of improved magnetometry by calculating the quantum Fisher information over a range of $N$ sufficient to draw reasonable conclusions about the scaling.

\begin{figure}[t]
\begin{center}
\includegraphics{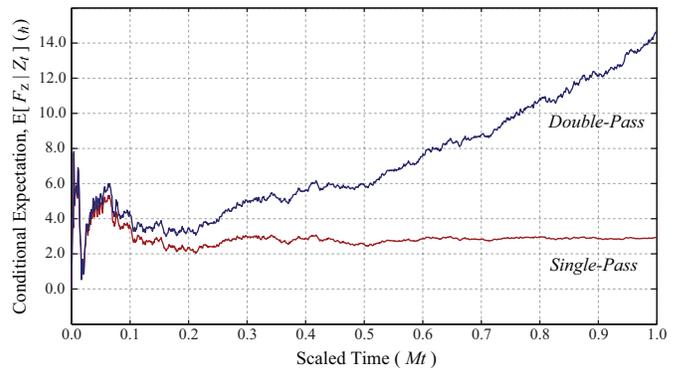}
\end{center}
\vspace{-6mm}
\caption{(color online) Comparison of the double-pass magnetometer depicted in Fig.\ (\ref{Figure::Schematic}) to one using a conventional single-pass Faraday measurement.   The simulated data was generated according to Eq.\ (\ref{eq:adjoint_quantum_filter}) with $B=0.1$ (units of $\gamma)$, $F=100$, $M=1.0$ and $K=1\times 10^{-4}$ (units of $1/\tau$).   The single-pass comparison was obtained by setting $K=0$.  \label{Figure::TrajectoryComparison}} 
\end{figure}

\vspace{.05in}
\noindent \textit{Effective Dynamics---} To estimate a static magnetic field oriented along the $y$-axis $\mathbf{B} = B \vec{y}$, the atomic sample is prepared into an $x$-polarized spin coherent state.   Like standard atomic magnetometers, a laser field with frequency $\omega_l$, far-detuned from the nearest atomic resonance $\omega_a$, is used to observe Larmor precession by measuring the $F_\mathrm{z}$ spin component via balanced polarimetry.   On its first pass, the incoming linearly-polarized optical field acquires a Faraday rotation proportional to $F_\mathrm{z}$.  Unlike standard Faraday spectroscopy, however, the probe laser passes through the atomic sample a second time  \cite{Sherson:2006a} prior to detection.  Just before the second pass, a quarter waveplate converts the first-pass Faraday rotation into elipticity, which the atoms perceive as a fictitious magnetic field during the second pass.  Since the second-pass field propagates parallel to the external magnetic field, the atoms see a total effective magnetic field that is the sum of $B$ and the $F_\mathrm{z}$-dependent elipticity.   As the atoms precess under $B$, the elipicity increases proportionately to $F_\mathrm{z}$ and further amplifies $F_\mathrm{z}$ (also the quantum uncertainty of the initial state), as illustrated by the simulation in Fig.\ (\ref{Figure::TrajectoryComparison}), which was generated using the model described below.

To analyze the proposed magnetometer quantitatively, we assume that the interaction between the atoms and the far-detuned optical probe is well-described by a vector polarizability \cite{Jessen:2004a}.  The collective atomic angular momentum  $\hat{\mathbf{F}}= ( \hat{F}_\mathrm{x}, \hat{F}_\mathrm{y}, \hat{F}_\mathrm{z})$ couples to the two polarization modes of the traveling wave probe, which act as a Schwinger-Bose field with Stokes opperators $\hat{\mathbf{s}} = ( \hat{s}_0, \hat{s}_\mathrm{x}, \hat{s}_\mathrm{y}, \hat{s}_\mathrm{z})$.  For convenience,  we choose the polarization basis given by defining the time-domain Schwinger boson annihilation operator 	$\hat{s}_t = \frac{1}{2} \int_{-\infty}^{+\infty}  g(\omega) \, \aDOmegaX \aHOmegaY e^{i \omega t} d \omega$ in terms of the annihilation operators $\aHOmegaX$ and $\aHOmegaY$ for the $x$ and $y$-polarzed plane-wave modes with frequency $\omega$ and form factor $g(\omega) \sim \sqrt{ \hbar \omega / 4 \pi c \epsilon_0}$.  The Stokes operators, $\hat{s}_\mathrm{x} = \hat{s} + \hat{s}^\dagger$ and $\hat{s}_\mathrm{y} = i( \hat{s} - \hat{s}^\dagger )$, are then reminiscent of quadrature operators and the interactions for each pass of the probe light through the sample can  be expressed as $\hat{H}^{(1)}  =   i \hbar \mu \hat{F}_\mathrm{z} ( \hat{s}_t^\dagger - \hat{s}_t )$ and $\hat{H}^{(2)}   =    - \hbar \kappa \hat{F}_\mathrm{y} ( \hat{s}_t^\dagger + \hat{s}_t )$ \cite{Chase:2009a}.

Developing a Markov approximation for the interactions, $\hat{H}^{(1)}$ and $\hat{H}^{(2)}$, is a standard problem in the theory of open quantum systems addressed by taking a weak-coupling limit \cite{Accardi:1990a,vanHandel:2005a} to obtain quantum It\^o equations 
\begin{eqnarray} \label{Equation::dUt1}
	d \hat{U}_t^{(1)} & = & \left\{ \sqrt{m} \hat{F}_\mathrm{z} ( d \hat{S}_t^\dagger - d \hat{S}_t )
		- \frac{1}{2} m \hat{F}_\mathrm{z}^2  dt \right\} \hat{U}_t^{(1)} \\
	d \hat{U}_t^{(2)} & = &  \left\{ i \sqrt{k} \hat{F}_\mathrm{y} (  d \hat{S}_t^\dagger + d \hat{S}_t )
		- \frac{1}{2} k \hat{F}_\mathrm{y}^2  dt \right\} \hat{U}_t^{(2)}  \,\,\,\,\,\,
		\label{Equation::dUt2}
\end{eqnarray}
for the first- and second-pass unitary propagators \cite{Chase:2009a,Sarma:2008a}.  In these expressions, $d \hat{S}_t^\dagger$ and $d \hat{S}_t$ are delta-correlated noise operators derived from the quantum Brownian motion $\hat{S}_t = \int_0^t \hat{s}_u du$.  The noise terms satisfy the quantum It\^o rules: $d \hat{S}_t d \hat{S}_t^\dagger = dt$ and $d \hat{S}_t^\dagger d \hat{S}_t = d \hat{S}_t^2 = (d \hat{S}_t^\dagger)^2 = 0$, and can be viewed heuristically as a consequence of vacuum fluctuations in the probe field.  The rates $m=2\pi |\mu g(\omega_l)|^2$ and $k=2\pi |\kappa g(\omega_l)|^2$ can be computed using the methods found for example in Refs.\ \cite{vanHandel:2005a}.  Assuming that the time-scale for the the light to propagate (twice) through the same is fast compared to the magnetic dynamics, the separate evolutions, Eqs.\ (\ref{Equation::dUt1}) and (\ref{Equation::dUt2}), can be combined into a single Markov limit, yielding the propagator for the double-passed dynamics
\begin{eqnarray}
	d \hat{U}_t & = & \left[ \left( i \sqrt{MK} \hat{F}_\mathrm{y} \hat{F}_\mathrm{z} - \frac{i}{\hbar}
		\hat{H}
		 - \frac{M}{2}  \hat{F}_\mathrm{z}^2   - \frac{K}{2}  \hat{F}_\mathrm{y}^2 
		 	\right) dt \right.  \\
		& & \left. \,\, +  \sqrt{M} \hat{F}_\mathrm{z} ( d \hat{S}_t^\dagger -
		d \hat{S}_t )+
		i \sqrt{K} \hat{F}_\mathrm{y} ( d \hat{S}_t^\dagger + d \hat{S}_t )
		 \right]  \hat{U}_t,  \nonumber
\end{eqnarray}
where $M=m/2$, $K=k/2$ and $\hat{H}= \hat{H}(B)$ is the Zeeman Hamiltonian from Eq.\ (\ref{Equation::HLarmor}) \cite{Chase:2009a}.

The measurement used to estimate $B$ is obtained by detecting the optical helicity of the probe laser, as this observable caries information about the $z$-component of the atomic spin.  A continuous measurement of the helicity is described by the time-evolved Stokes operator $\hat{Z}_t =  \hat{U}_t^\dagger (  \hat{S}_t^\dagger + \hat{S}_t )  \hat{U}_t$, with the dynamics $d \hat{Z}_t = 2  \sqrt{M}  \hat{U}^\dagger_t  \hat{F}_\mathrm{z}\hat{U}_t  \, dt + ( d \hat{S}_t^\dagger + d \hat{S}_t )$.  It can be shown that the observations constitute a classical stochastic process $Z_t$ \cite{Belavkin:1999a,vanHandel:2005a}.   The techniques of quantum filtering theory \cite{Belavkin:1999a,Wiseman:1993a,vanHandel:2005a} can then be used to obtain the best least-squares estimate of the expectation value of any atomic operator $\hat{X}$ conditioned on the measurement $\pi_t(\hat{X}) = \mathbbm{E}[ \hat{X} | Z_t ] = \mathrm{tr}[ \hat{X}\hat{\rho}_t]$ where the conditional density operator satisfies 
\begin{eqnarray} 
    d \hat{\rho}_t & = & i\gamma B[\hat{F}_\mathrm{y},\hat{\rho}_t]dt 
    	+ \sqrt{KM}[\hat{F}_\mathrm{y},\{\hat{F}_\mathrm{z},\hat{\rho}_t\}]dt \nonumber \\
	&&    + M \mathcal{D}[\hat{F}_\mathrm{z}] \hat{\rho}_t dt 
             + K\mathcal{D}[ \hat{F}_\mathrm{y}] \hat{\rho}_t dt \label{eq:adjoint_quantum_filter} \\
         &&
             + (\sqrt{M}\mathcal{M}[\hat{F}_\mathrm{z}] \hat{\rho}_t 
             + i\sqrt{K}[\hat{F}_\mathrm{y}, \hat{\rho}_t]) dW_t. \nonumber
\end{eqnarray}
Here, the innovations $dW_t = dZ_t - 2\sqrt{M}\mathrm{tr}[\Fz\rho_t]dt$ are a Wiener process, $\mathbbm{E}[dW_t] = 0$, $dW_t^2 = dt$, and the superoperators are defined as $\mathcal{D}[\hat{X}]\hat{\rho}_t = \hat{X} \hat{\rho}_t \hat{X}^\dagger- (\hat{X}^\dagger \hat{X} \hat{\rho}_t +\hat{\rho}_t \hat{X}^\dagger\hat{X})/2$ and $\mathcal{M}[\hat{X}]\hat{\rho}_t =\hat{X} \hat{\rho}_t + \hat{\rho}_t \hat{X} - \mathrm{tr}[ (\hat{X}+\hat{X}^\dagger) \rho_t] \hat{\rho}_t$.  A derivation of Eq.\ (\ref{eq:adjoint_quantum_filter}) can be found in Ref.\ \cite{Chase:2009a}.

\vspace{0.1in}
\noindent\textit{Quantum Fisher Information Results---} The quantum filter Eq.\ (\ref{eq:adjoint_quantum_filter}) can be used to simulate individual measurement realizations of the double-pass magnetometer [c.f., Fig.\ (\ref{Figure::TrajectoryComparison})].  Furthermore, one can evaluate the quantum Fisher information (at least numerically) via a finite-difference approximation to the logarithmic derivative\begin{equation}
	 \hat{\mathfrak{L}}_t(B) = 2 \frac{\partial \hat{\rho}_t(B)}{ \partial B} 
	 	\approx \frac{\hat{\rho}_t(B+dB) - \hat{\rho}_t(B-dB)}{dB}
\end{equation}
by evolving $\hat{\rho}_t(B)$, $\hat{\rho}_t(B+dB)$ and $\hat{\rho}_t(B-dB)$ under the same noise realization for $dB \ll 1$ \cite{Chase:2009a}.   On an individual trajectory basis, the Fisher information calculated using the conditional density operator $\mathcal{I}_t | Z_t = \mathrm{tr}[ \hat{\mathfrak{L}}_t^2 \hat{\rho}_t(B)]$ is also conditioned on the particular measurement realization that generated $\hat{\rho}_t$ and must be averaged over many realizations to obtain the unconditional quantum Fisher information $\mathcal{I}_t = \mathbbm{E}[ \mathcal{I}_t | Z_t ]$.   The lower bound $\delta\tilde{B}_\tau$ can then be obtained from Eq.\ (\ref{Equation::CRError}) with statistical errorbars given by $\sigma (\delta\tilde{B}_\tau) = \mathcal{I}_\tau^{-3/2} \sigma[ \mathcal{I}_\tau | Z_t ] / 2$.

\begin{figure}[t]
\begin{center}
\includegraphics{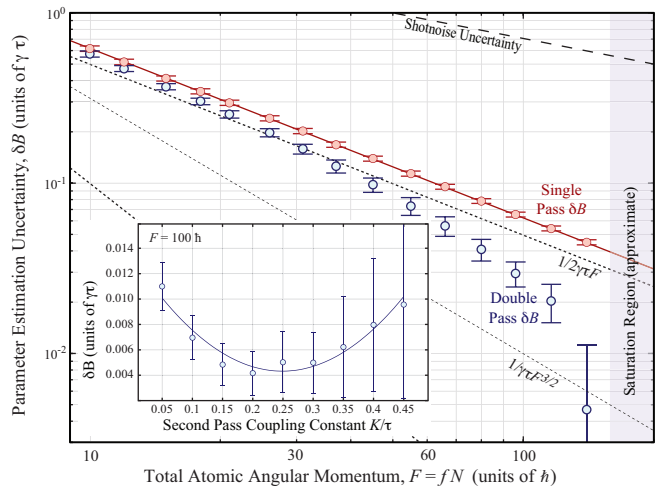}
\end{center}
\vspace{-5mm}
\caption{(color online) Comparison of the estimation uncertainty $\Delta \tilde{B}$ as a function of the total atomic angular momentum (proportional to $N$) for double-pass and single-pass atomic magnetometers determined by calculating the quantum Fisher Information with $M=1$ (in units of $1/\tau$) and $K=1\times10^{-4}$ chosen to be optimal for $F=140 \hbar$ . \label{Figure::StrongFScaling}} 
\end{figure}

We calculated $\mathcal{I}_t(B)$ over a range of spin quantum quantum numbers $F=Nf$ spanning more than an order of magnitude to determine a lower bound on the magnetic field estimation uncertainty using  Eq.\ (\ref{Equation::CRError}).   Our results indicate that the Fisher information depends heavily upon the choice of the coupling strengths $M$ and $K$, which is not surprising since the measurement strength $M$ determines how much spin-squeezing is generated and $K$ determines the strength of the effective nonlinearity.   Like any measurement procedure that involves amplification, both the signal and noise are affected, and optimal performance requires choosing the correct gain.

If one choses $M = 1 / \tau$, to obtain an optimal spin-squeezed state at the final time $t=\tau$ \cite{Geremia:2003a}, then it is straightforward to optimize over the nonlinearity $K$, as illustrated in the inset of Fig.\ (\ref{Figure::StrongFScaling}) for $F=100 \hbar$.  We found that the optimal value $K^*$ depends upon the number of atoms, and that the Fisher information saturates and then decreases if the number of atoms exceeds the value of $N=F_\mathrm{sat}/f$ used to compute $K^*(F)$.    Figure \ref{Figure::StrongFScaling}) shows the behavior of $\delta \tilde{B}_\tau$ as a function of $F$ up to the saturation point $F < F_\mathrm{sat}\sim 150$.  The largest value of $F$ prior to saturation yields a $\delta\tilde{B}_\tau$ that is slightly below the bound $1/\tau \gamma F^{3/2}$ that would be obtained for a two-body coupling Hamiltonian and an initially separable state $\hat{\rho}_0$ \cite{Boixo:2008a}.  Despite this saturation of the quantum Fisher information for $F > F_\mathrm{sat}$ at a given choice of $K$, one can choose the value of $K^*$ such that saturation occurs only for $F_\mathrm{sat} > F_\mathrm{max}$ over any specified finite range $F \le F_\mathrm{max}$.  An improvement beyond $1/N$ scaling can be achieved over any physically realistic number of particles.

\begin{figure}[t]
\begin{center}
\includegraphics{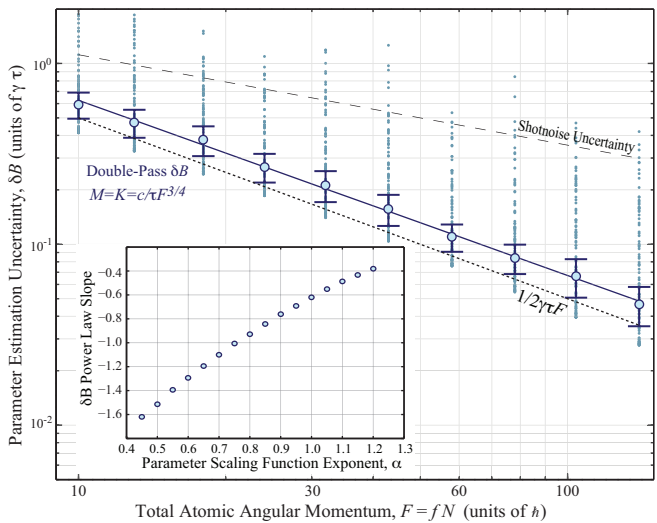}
\end{center}
\vspace{-5mm}
\caption{(color online) Evidence that the field estimation uncertainty $\Delta \tilde{B}$ can be made to scale as a power law $\delta\tilde{B}_\tau \sim 1/N^k$ by decreasing the parameters $M$ and $K$ as a function of the total angular momentum $F$ according to Eq.\ (\ref{Equation::MKScaling}) with $\alpha \approx 3/4$.  The power-law fit (solid line) has a slope of $-0.97$. \label{Figure::WeakFScaling}} 
\end{figure}

A second approach to avoiding saturation of the Fisher information for large $F$ is to scale the parameters $M$ and $K$ as a decreasing function of $F$.   For practical considerations, it is also desirable to set $M=K$ as these parameters are determined by the atom-field coupling strengths on the first and second pass interactions, thus quantities such as the laser intensity and detuning not easily changed between the two passes.  We have found that scaling $M$ and $K$ according to the functional form
\begin{equation} \label{Equation::MKScaling}
	M = K = c / \tau F^\alpha, 
\end{equation}
where $c$ and $\alpha$ are constants, leads to a power-law scaling for the uncertainty bound $\delta \tilde{B}_\tau \sim 1/N^k$.  The inset plot in Fig.\ (\ref{Figure::WeakFScaling}) shows the slope of a linear fit of $\log_{10}\delta\tilde{B}_\tau$ to $\log_{10}F$ (i.e., a slope of $k=-1$ corresponds to the Heisenberg uncertainty scaling) as a function of $\alpha$ (with $c$ chosen so as to avoid the saturation behavior described above).  As demonstrated by the data points in Fig.\ (\ref{Figure::WeakFScaling}), it is possible to achieve $1/N$ scaling (to within a small prefactor offset) with $\alpha = 0.77$ and $c=0.589$.   The distribution of conditional uncertainties $\delta\tilde{B}_\tau | Z_t$ for the statistical ensemble of measurement realizations [dots in Fig.\ (\ref{Figure::WeakFScaling})] is depicted for the different values of $F$.  The mean and uncertainty of this distribution are denoted by the circles and errorbars, and a fit to this data gives $\delta\tilde{B}_\tau \sim F^{-0.97}$.

Extrapolating the values of $M$ and $K$ that would be required to achieve a sensitivity $\delta\tilde{B}_\tau$ corresponding to the Heisenberg uncertainty scaling Eq.\ (\ref{Equation::DeltaBHL}) with $N \sim 10^8$ (small for typical experiments) implies that one would only require $M\tau \sim 10^{-7}$.  Since $M\tau = 1$ corresponds to a maximally spin-squeezed state at time $t=\tau$, these results suggest that $1/N$ scaling can be achieved by an extremely weak measurement, with no appreciable generation of conditional spin-squeezing.   These results suggest that it may be much easier than previously believed to to outperform the shotnoise scaling since doing so should not require generating an entangled state of the atoms.

\vspace{.1in}
\noindent\textit{Conclusion---} Our results highlight that there are two complementary ways to improve metrological sensitivity: (1) reducing the quantum noise of the probe; and (2) enhancing the accumulated phase acquired by the probe due to its interaction with the metrological quantity.   Our procedure does both--- probe uncertainty is reduced by conditional spin squeezing, and phase accumulation is enhanced by coherent optical feedback.   It does, however, remain an open question to reconcile precisely how the effective dynamics observed here can be understood using the Hamiltonian framework of Ref.\ \cite{Boixo:2007a}.   This outstanding problem is extremely important since effective multibody interactions, like those developed here, are likely to be essential to optimizing sensitivity in precision measurements.  This work was supported by the NSF (PHY-0639994) and the AFOSR (FA9550-06-01-0178).  Please visit http://qmc.phys.unm.edu/ to download the simulation code used to generate our results as well as all data files used to generate the figures in this paper.

\end{document}